\documentclass{stavanger}

\usepackage[utf8]{inputenc}
\usepackage{amssymb,amsmath,amsfonts,amsthm}
\usepackage{hyperref}
\usepackage{natbib}
\usepackage{graphicx}
\usepackage{tikz}
\usepackage{cleveref}
\usepackage{ulem} 
\usepackage{placeins}

\startlocaldefs

\endlocaldefs

\begin{document}


\begin{frontmatter}

\begin{fmbox}


\dochead{Research Article }{FP}


\title{Density of observables from local derivatives}


\author[
]{\inits{FL}\fnm{Rasmus N.} \snm{Larsen}}


\address[id=aff1]{
  \orgname{Faculty of Science and Technology}, 	 
  \street{University of Stavanger},                     		 
  \postcode{4021}                               			 
  \city{Stavanger},                              				 
  \cny{Norway}                                   				 
}



\end{fmbox}


\begin{abstractbox}

\begin{abstract} 
We derive a formula to calculate the local change to the log of any density of states for smooth real observables. Using this in Monte-Carlo simulations, we are able to calculate the expectation value of the observable with a precision often better than standard sampling. The method can be applied to previously generated configurations, as long as the analysis uses the same action used to generate the configurations. We show that for observables such as Wilson line correlators, errors are reduced by up to 4 times.
\end{abstract}



\end{abstractbox}


\end{frontmatter}


\section{Density of States}
In this paper, we wish to calculate the density of states, which is a measure of how often a distribution takes on certain values, like a velocity distribution or the variation in the action or energy. We will focus on systems that can be written as
\begin{eqnarray}
\langle f(O) \rangle &=& \frac{\int d^D x f( O(x) ) \exp(-S(x))}{\int d^D x  \exp(-S(x))}
\end{eqnarray}
where $x$ is a D dimensional space of some variable we wish to integrate over and S and O are real functions that are differentiable at all $x$, excluding a finite amount of points in $x$ and f can be any function. We will refer to O as the observable and S as the action. We wish to rewrite the system as
\begin{eqnarray}
\langle f(O) \rangle &=& \frac{\int dO  f(O) \rho (O)}{\int  dO \rho (O)}
\end{eqnarray}
where $\rho (O)$ is the density of states for the observable $O$, weighted by $\exp(-S(x))$, also known as the probability of $O$. 

There are several ways to obtain such a distribution. One can for instance sample with $\exp(-S(x))$ as the probability distribution and then histogram the obtained values of $O$. A more sophisticated solution is to look for a local $\rho(O) =\exp(a(O))$ around several values of O such that $a(O)$ cancels the slope in the density, which is seen by getting a symmetric sample around $O$  \cite{Langfeld:2012ah}, often referred to as the LLR method due to the last names of the authors.

Reconstructing a density of states from Monte-Carlo simulations often produces behavior as seen in figure \ref{fig:x4hist} which shows an almost Gaussian distribution. The precision of the average 
\begin{eqnarray}
\langle f(O) \rangle = \frac{\sum _n ^{N} f(O(x_n))}{N}
\end{eqnarray}
behaves as the square root of the number of measurements, but the slope of the Gaussian is seen to fluctuate a lot, even for high statistics. We will refer to the above equation as the standard way to obtain an average expectation value. The most typical choice for $f$ is $f(O)=O$, which gives the average of the observable itself. Another choice for $f$ is $f(O) = \exp(iO)$ which is interesting for studying sign problems.

 We want to use the additional information about smoothness to improve the predictions for different observables. 
This is similar to what is done in the LLR method \cite{Langfeld:2012ah}, but instead of finding the relative change from finding the slope that makes the sampling constant in a small window, we will instead try to calculate the slope from derivatives of the observable $O$.

In \cite{Larsen:2022pni} it was shown that the full integral can be expressed in terms of line integrals that never cross, but instead have the relative volume between the lines increase or decrease, as they get closer or further away from other lines, similar to integration using radial coordinates which require a factor of $r^{D-1}$ for the radial part. In the paper, it was used to calculate partial contributions to an integral by looking at the change along the imaginary part of the action. Here we will instead look at an infinitesimal change along these lines, and we will be looking at the change along the observable. This does require the observable to be smooth at all but a finite number of points. 

The main observation for this approach is that for any $D$ dimensional vector, there exist $D-1$ orthogonal vectors. If the first vector is pointing in the direction of change of some observable, then all the other $D-1$ directions will be constant in the observable. We therefore only need to find how the volume  changes along the one direction where the observable is changing, in order to understand if more or less states with a larger or smaller value compared to the current value of $O$ exist. If one does this at all possible values of x, along an observable O, one will have found how the probability distribution of O, which is the density of states, will be changing.

We consider a small volume around a point in the integration space $x_i$, where i goes from $1$ to $D$ and indicates the different components of x. We then change $x_i$ along the change to the observable, which we will call $O$, by following $\frac{\partial O}{\partial x_i} \equiv O_i$, until the observable has changed by $\epsilon$, where $\epsilon$ is an infinitesimal change.  In this paper, we will use the notation that a subscript on the observable O is a partial derivative with respect to the variable $x_i$ and summation when a subscript appears twice or to a power. In case O is complex, the variable needs to be split into 2 parts, the real and imaginary part, which will be treated separately, i.e. the observable needs to be real. 
 Making an infinitesimal change $x_i \to x_i + dx_i = x_i+ \epsilon O_i (x)$ gives

\begin{eqnarray}
O(x_i+\epsilon O_i) &=& O(x_i)+\epsilon O_i^2+R(\epsilon ^2) 
\end{eqnarray}
where $R(\epsilon ^2) $ is the remaining part of the expansion with $\epsilon ^2 $ or higher (R is often called $\mathcal{O}$, but to avoid confusion with observable we use R in this paper instead).
We observe that for the change $dx_i$  to correspond to a change to $O$ of size $\epsilon$, we need to normalize $dx_i$ to $dx_i = \epsilon\frac{O_i}{O_j^2}$.

We then need to calculate what change has happened to the small volume around $x_i$ by the change $d x_i$. If we do this for all $x_i$ with a specific value of O, i.e. O(x)=O, we will then obtain how much larger the volume at $O'=O+\epsilon$ is compared to $O$. This is equivalent to
\begin{eqnarray}
\frac{1}{\epsilon}\log(V(O+\epsilon)/V(O)) &=& \frac{1}{\epsilon}\log(\rho(O+\epsilon)/\rho(O)) = \frac{d \log(\rho(O))}{d O} +R(\epsilon)
\end{eqnarray}

If we can calculate this change in the volume, it gives us access to the derivative of the log of the density of states.  

\begin{figure}
\centering
\includegraphics[width=0.9\textwidth]{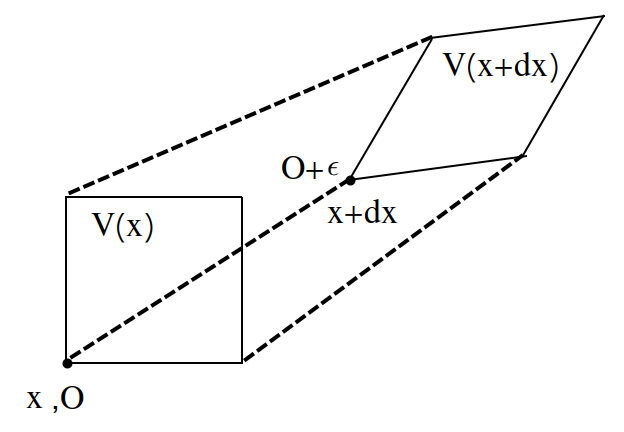}
\caption{2d Sketch of how the local volume changes under the change $x \to x + dx$. The change $dx_i=\epsilon\frac{O_i}{O_j ^2}$ makes a change of $\epsilon$ to the observable O. The change to the volume comes from the $x$ dependence of $dx$ such that the corners of the box evolves differently from $x$. Adding all such changes to the volume up for all $x$ when $O(x)=O$ gives the relative change in volume between $O$ and $O+\epsilon$.}
\label{fig:Vsketch}
\end{figure}

Following the ideas from equation 8 from \cite{Larsen:2022pni}, we calculate the relative change to the local volume around point $x_i$ from the change to the space spanned by D vectors $v_{ij}$ around $x_i$ evolved by $\epsilon\frac{O_i}{O_j^2}$. We sketch the method in figure \ref{fig:Vsketch}.
\begin{eqnarray} 
\left[ dx_i(x_u+\epsilon _2 v_{up})-dx_i(x_u)\right]/\epsilon_2 &=& \epsilon  \frac{\partial }{\partial x_j}\left(\frac{O_i}{O_k ^2} \right) v_{jp}  +R(\epsilon_2) \nonumber \\
\frac{V(O + \epsilon)}{V(O)} = \det\left[I_{ij}+\epsilon  \frac{\partial }{\partial x_j}\left(\frac{O_i}{O_k ^2} \right)\right] &=& 1+\epsilon   \frac{\partial }{\partial x_j}\left(\frac{O_j}{O_k ^2} \right)+ R(\epsilon ^2) 
\end{eqnarray}
In the second line we have chosen the vector space to be the identity matrix, but since $\det(v)$ can be factorized out, the relative change does not depend on the orientation of the vector space, unlike complex evolution. The change to the relative volume is therefore given by the log of the above result, which we define as $dV$
\begin{eqnarray}
 dV &\equiv & \partial _{x_i} (\frac{O_i}{O_j^2})  =  \frac{O_{i,i}}{O_j^2}-\frac{2O_i O _{i,j} O_j}{(O_k^2)^2}
\end{eqnarray}
If we were only interested in the observable's random distribution, this would be the final formula. We are however interested in systems that have been sampled with a weight of $\exp(-S(x))$. The relative change to the density of states $\rho(O)$ will therefore be given by

\begin{eqnarray}
V(O+\epsilon) &=& V_{norm}(\sum _{O(x)=O}  \exp[\epsilon dV -(S(x_i+dx_i)-S(x_i))])  \nonumber \\
 &=& V_{norm}N_{samples}(1+\epsilon (\langle dV\rangle -\langle \frac{S_i O_i}{O_j ^2}\rangle )+R(\epsilon ^2))\\
\log[V(O+\epsilon)/V(O)] &=& \log[ 1+\epsilon (\langle dV\rangle -\langle \frac{S_i O_i}{O_j ^2}\rangle )]+R(\epsilon ^2)) ) \\
&=& \epsilon (\langle dV\rangle  -\langle \frac{S_i O_i}{O_j ^2}\rangle )+R(\epsilon ^2) \nonumber \\
d _O \log (\rho (O)) &=&  \langle dV\rangle -\langle \frac{S_i O_i}{O_j ^2}\rangle   \\
                &=&  \langle \frac{O_{i,i}}{O_j^2}-\frac{2O_i O _{i,j} O_j}{(O_k^2)^2} -\frac{S_i O_i}{O_j ^2}\rangle  \label{eq:dO}
\end{eqnarray}
where we have assumed that every $x_i$ was sampled in a Monte-Carlo process with $S$ as the probability, such that the initial volume is set to $V_{norm}$ and $\exp[\epsilon dV -(S(x_i+dx_i)-S(x_i))]$ gives the relative change. In case $S=O$ the sampling does not have to have been done with $\exp(-S)$, as all x for $O(x)=O$ will have the same weight of $\exp(-S)$. The expectation value $\langle \rangle$ is here with respect to $O(x)=O$.

The  term $\langle dV\rangle $ is the relative change to the volume calculated at all points with a specific value of O. The term $\langle \frac{S_i O_i}{O_j ^2}\rangle$ comes from the change to the action when the observable is changed, which is the change to the action when $x_i$ is changed by $dx_i$. We therefore need to calculate the trace of the double derivative of the observable and the derivative of $O$, $O_j^2$ and $S$. 

\section{Simple Examples}

Let's take a look at how eq. (\ref{eq:dO}) works in a simple case that can be solved analytically. For the observable $O=S=x_i^2$ for a D dimensional space, we have that the action correction term becomes -1. This is always true for $S=O$. For the other terms we have

\begin{eqnarray}
O_i &=& \partial_ {x_i} (x_j^2) = 2x_i \\
O_{i,i} &=& \partial ^2_{ x_i} (x_j^2) = 2D \\
O_i^2 &=& 4x_i ^2 = 4S \\
O_i O_{i,j} O_j &=& 2 x_i 2 \delta _{i,j} 2x_j = 8x_i^2 = 8S \\
d_S \log (\rho(S)) &=& \frac{2D}{4S} -\frac{2*8S}{(4S)^2}-1 = \frac{1}{S}(D/2-1)-1
\end{eqnarray}
The last differential equation can then be solved, which gives
\begin{eqnarray}
\rho(S) &=& \rm{ Const}*\exp(\log(S)(D/2-1)-S) = \rm{ Const} * S^{D/2-1}\exp(-S) 
\end{eqnarray}
which can also be obtained by going to radial coordinates. The constant can be fixed by $\int \rho(S) d S = 1$. Generally, however, we cannot simply go to radial coordinates, and the derivatives of O with respect to x cannot be expressed only in terms of S. In those cases, we need to evaluate the expectation values from computer simulations.

\begin{figure}[h!]
\centering
\includegraphics[width=0.9\textwidth]{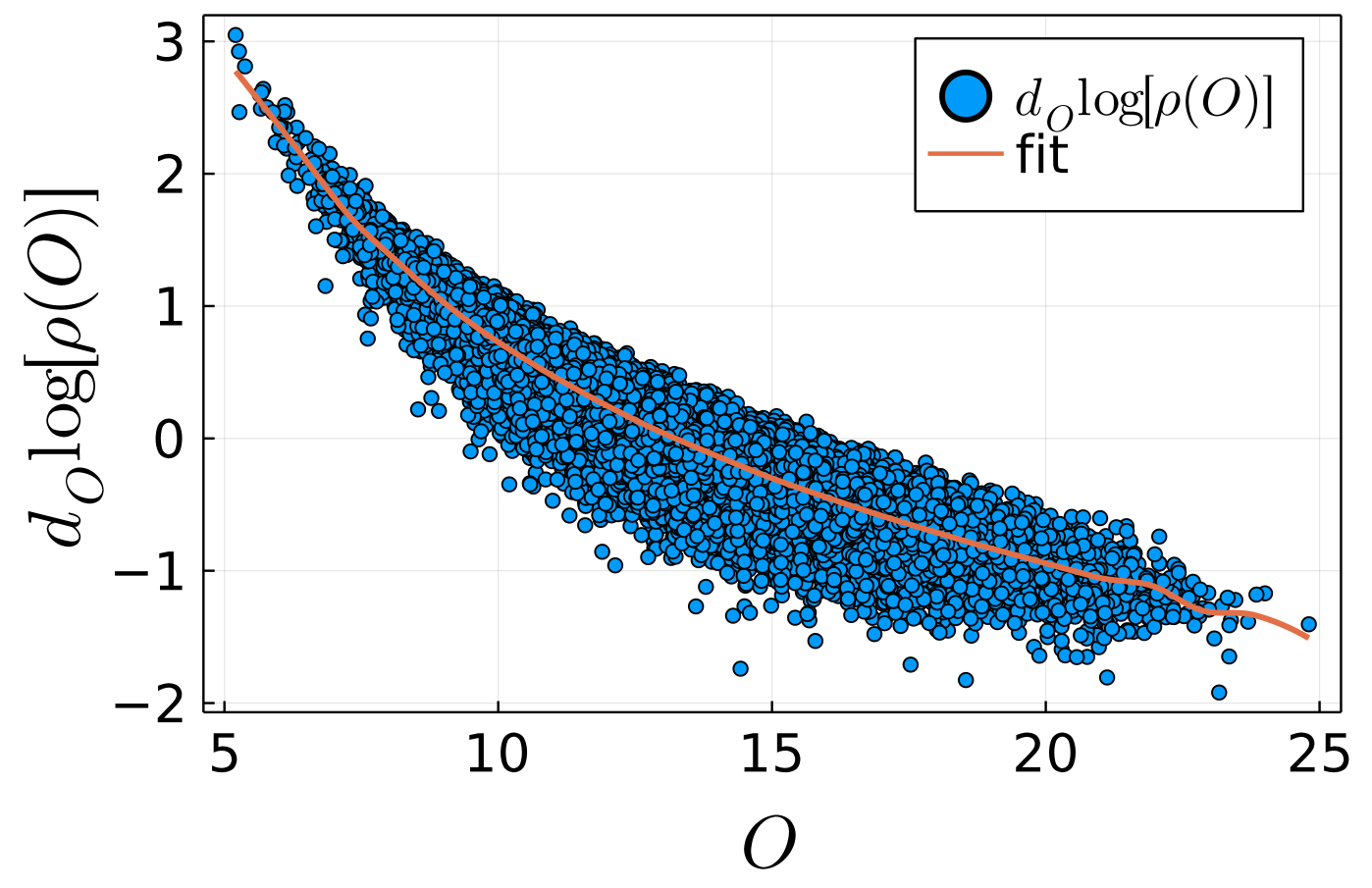}
\caption{(blue points) $10^5$ samples for $d_O \log (\rho(O))$ and (orange line) a 20 order spline fitted to the data for $D=40$, $O=x_i ^2$ and $S=x_i ^4$.}
\label{fig:x4dl}
\end{figure}

We first try to use it on the same observable $O=x_i^2$, but with the action $S=x_i^4$. In one dimension the procedure is trivial, but for higher dimensions the action and its derivative can take many values for a specific $O$. We see in figure \ref{fig:x4dl} the obtained distribution for $d_O \log (\rho(O))$ for a 40 dimensional system. The example can be calculated exactly as
\begin{eqnarray}
\langle x_i^2 \rangle &=& \frac{\int x_i^2 \exp(-x_i^4) d^{40} x}{\int \exp(-x_i^4) d^{40} x} = 40 \frac{\int x^2 \exp(-x^4) d x}{\int \exp(-x^4) d x} = 13.5196
\end{eqnarray}

\begin{figure}[h!]
\centering
\includegraphics[width=0.9\textwidth]{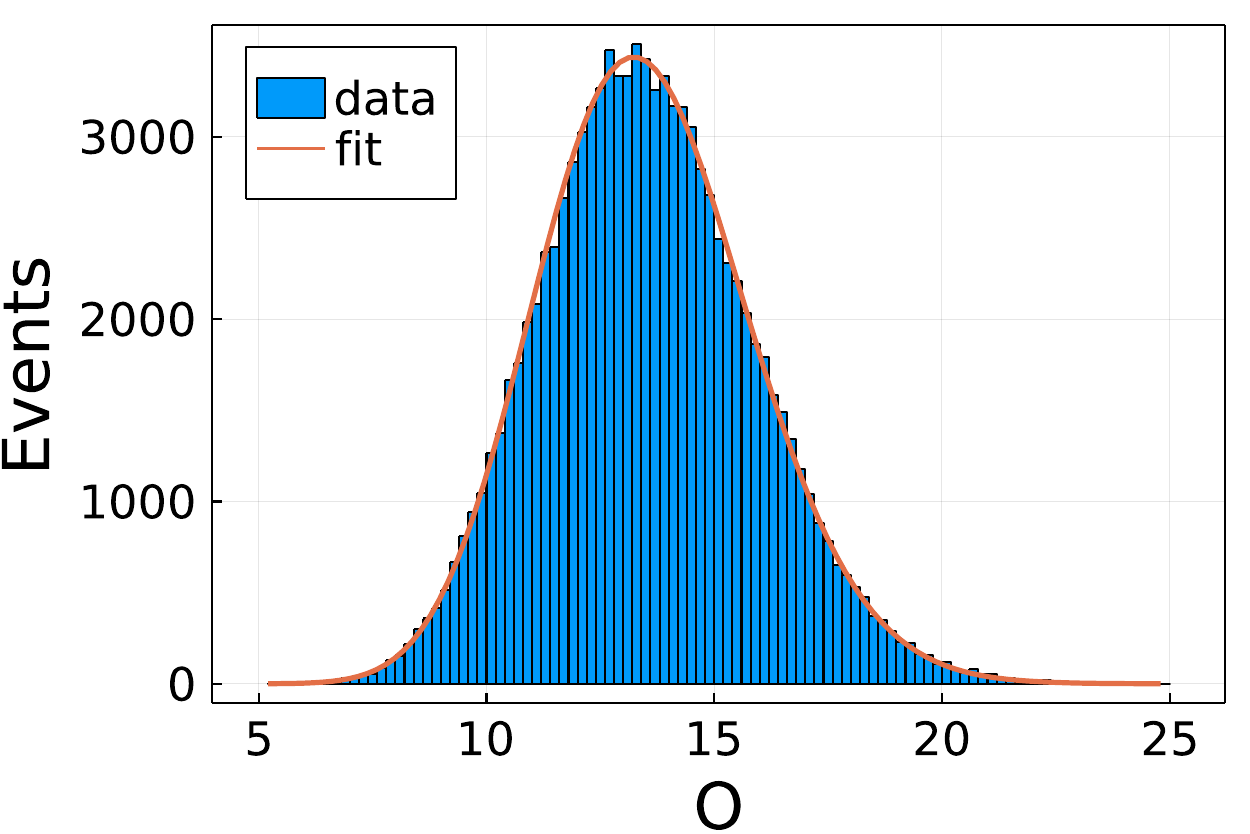}
\caption{The histogram of $10^5$ measurements for a HMC sampling of $O=x_i^2$ for $D=40$ for $S=x_i ^4$. The line plotted on top is the density of states $\rho(O)$ calculated from the fit to $d_O \log (\rho(O))$.}
\label{fig:x4hist}
\end{figure}

 In a numerical simulation, there are (within machine precision) never 2 values of $O(x)$ that give the same value. We therefore cannot do an average over $O(x)=O$, instead to obtain $\rho (O)$ we have fitted with splines of order 10-30 and chosen the one with the smallest error bars. Results are shown in table \ref{table:x4} for the expectation value of $O$ and in figure \ref{fig:x4hist} for the obtained density of states $\rho(O)$.

\begin{table}[h!]
\begin{center}
\begin{tabular}{||c c c  ||} 
 \hline
N & Observable & Result   \\ [0.5ex] 
 \hline\hline
 $10^5$ & O & $13.515 \pm 0.018$  \\ 
 \hline
 $10^6$ & O & $13.512 \pm 0.005$  \\ 
 \hline  
  $10^7$ & O & $13.521 \pm 0.001 $ \\ 
 \hline  
  $10^5$ & dV & $13.523 \pm 0.007 $ \\ 
 \hline 
\end{tabular}
\caption{Expectation value of $x_i^2$ for $S=x_i^4$ for standard sampling method O and for the density of states from derivative dV. The exact result is $13.5196$.}
\label{table:x4}
\end{center}
\end{table}

\section{SU3}
Many lattice simulations use group elements of SU3 as the variable. In this case, we have to consider if the change to the group element corresponds to a trivial metric. To find if the SU3 group introduces new changes to the volume, we therefore have to calculate \cite{latticebook}
\begin{eqnarray}
g_{kl} &=& 2 tr\left(\frac{\partial U}{\partial \alpha _k}\frac{ \partial U ^ \dagger}{\partial \alpha _l} \right) \label{eq:gkl}
\end{eqnarray}
where $\alpha _k$ are the coordinates of SU3 and the Haar measure is given by $dU=\nu \sqrt{det(g)} \Pi _k d\alpha _k$. $d\alpha _k$ is the change to the local coordinates and $\nu$ is a normalization constant.
We define a change to SU3 as multiplication from the left

\begin{eqnarray}
\Delta U &=& \exp(i\epsilon _k \tau _k)U  -U
\end{eqnarray}

where $i$ is the imaginary unit, and we will therefore use $j$, $k$ and $l$ for the index of $x$ from now on. Plugging this into equation (\ref{eq:gkl}), we find 
 \begin{eqnarray}
 2 tr\left(\frac{\partial U}{\partial \alpha _k}\frac{ \partial U ^ \dagger}{\partial \alpha _l} \right) &=&  2 tr[(i\tau _k) U (i \tau _l U) ^\dagger] = 2 tr(\tau _k \tau _l) = 4\delta _{k,l}
\end{eqnarray}  
 such that the metric under these changes is flat, and therefore does not introduce any new volume change. The derivation for $d _O \log (\rho (O))$ in eq. (\ref{eq:dO}) should therefore still hold true.

We calculate $O_j$ as $O(x+\epsilon) = O(\exp(i \epsilon _j \tau_j) U) = O+\epsilon_j O_j +R(\epsilon^2)$. It is often easier to calculate $O_k O_{k,j} O_j$ as the change to $O_k^2$ along the change $dx_j$ by a finite numerical difference $2 O_k O_{k,j} O_j = \frac{1}{\epsilon}( O_k(\exp(i \epsilon O_j \tau_j) U)^2-O_k(\exp(-i \epsilon O_j \tau_j) U)^2)+R(\epsilon ^2)$ to avoid having to calculate a full matrix.

To test out the method in SU3, we look for the observable Re[Tr(U)] for an 8 parameter system, for the 8 generators. The term $O_k O_{k,j} O_j$ has been calculated with both the exact formula and finite difference. For double precision, this is in agreement at order $10^{-9}$ when $\epsilon = 10 ^{-5}$. The action is chosen to be $S= (Re[Tr(U)]-1) ^2$. The results are shown in table \ref{table:su3}.

\begin{table}[h!]
\begin{center}
\begin{tabular}{||c c c  ||} 
 \hline
N & Observable & Re  \\ [0.5ex] 
 \hline\hline
 $10^5$ & O & $0.5313 \pm 0.0094 $  \\ 
 \hline
 $10^6$ & O & $0.5194 \pm 0.0037 $  \\ 
 \hline  
  $10^5$ & dV & $0.5244 \pm 0.0002 $  \\ 
 \hline 
   $10^6$ & dV & $0.5242 \pm 0.00009 $  \\ 
 \hline 
\end{tabular}
\caption{Results for the expectation value of $O= Re[Tr(U)]$ for $S= (Re[Tr(U)]-1) ^2$ calculated from standard sampling  $O$ and the density method $dV$ introduced in this paper.  Results depend slightly on bin size for fit, probably due to the pole at $O=-1$.}
\label{table:su3}
\end{center}
\end{table}

\begin{figure}
\centering
\includegraphics[width=0.9\textwidth]{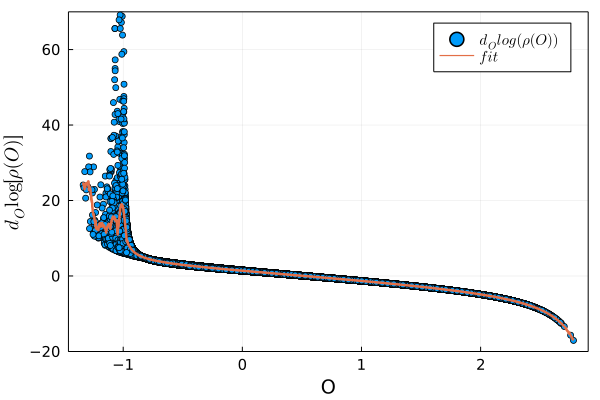}
\caption{(blue points) $10^6$ samples for $d_O \log (\rho(O))$ for observable $Re[Tr(U)]$ for $S=(Re[Tr(U)]-1)^2$ and (orange line) fit done with 2. order polynomial that fitted in the range $1/100$ of the full range around each point.}
\label{fig:su3dl}
\end{figure}

We see in figure \ref{fig:su3dl} that besides the endpoints $O=\frac{-3}{2},3$ where the density of states $\rho(O)$ goes to zero ($d_O \log (\rho(O))$ goes to plus/minus infinity), there is also a pole around $O=-1$. The pole structure seems to be of size $|(O+1)|^{-1/6}$ from numerical estimation, which means that the contribution from the pole is finite. This means that fitting with a smooth function around $O=-1$ will not work. Instead, the derivative of the log of the density was found by fitting with a 2. order polynomial in the range $1/100$ of the full range around each point.


\section{Wilson line correlator}
The Wilson line correlator is defined as 
\begin{eqnarray}
C(\tau,r) &=& \sum _{\tau ' ,x} Tr[(\Pi _{j=0} ^{\tau / a-1} U_4(\tau '+aj,x))(\Pi _{j=0} ^{\tau /a-1} U_4(\tau '+aj,x+r))^\dagger]
\end{eqnarray}

where the multiplication starts to the left and then moves to the right. The sum is over all possible starting locations in 4d and $a$ is the lattice spacing. As can be seen from the above equation, the Wilson line correlator is not gauge invariant. We therefore have to fix the gauge. We use coulomb gauge. It is possible that the derivative $O_i$ will take the configuration out of the gauge fixed state, thus requiring a derivative on the gauge fixing. We have however previously \cite{Bazavov:2023dci} observed that small variations in the precision of the gauge fixing are not important for the Wilson line correlator, especially at late complex time $\tau$, and we therefore ignore this possible effect for now.  

The derivative $O_j$ when $O=C$ can then be calculated by inserting a $i \tau_j$ in front of $U_4(x)$. Due to the sum over the entire volume, we can write 
\begin{eqnarray}
C(\tau,r) = C(\tau,-r) ^\dagger
\end{eqnarray}
which means that $C$ is real when averaging over all r of the same length.
From this, we can see that we only need to change the links without $\dagger$ at x for $O_j(x)$ if we add the corrections from $r$ and $-r$, as the real part will be the same for the contribution from chaining the $U_4 ^\dagger$ contributions in $C$, as long as we remember a factor of 2. It was found that the fastest way to implement the derivative of the Wilson line correlator was to sum over all r of the same length $|r|$ before taking the derivative to find $O_j$. This is because the part containing the Wilson line moving forward in complex time $\tau$ (the part containing $U_4$ but not $U_4 ^\dagger $) can be factorized out of the sum over $r$ 

\begin{eqnarray}
C(\tau,r) &=& \sum _{\tau ' ,x} Tr[(\Pi _{j=0} ^{\tau /a-1} U_4(\tau '+aj,x)) \times \nonumber \\
     &&\sum _{r=|r|} (\Pi _{j=0} ^{\tau /a-1} U_4(\tau '+aj,x+r))^\dagger] /(\sum _{\tau ',x,r=|r|} 1)
\end{eqnarray}

The double derivative becomes $O_{j,j} = -(16/3)2\tau O$ since $\exp(i\epsilon _j\tau_j) = 1+ i\epsilon_j \tau_j -(\epsilon _j\tau _j)^2/2$ and we use that $\tau_j ^2 = 16/3$ for the used representation of su3. We get this contribution for every link in the product, which is $2\tau$. This contribution is counted once for each sum over separations $r$ that has length $|r|$.

$\partial _{x_j} S$ is taken from the Hybrid-Monte-Carlo simulation code from which the configurations were generated. In this case, it is SIMULATeQCD \cite{HotQCD:2023ghu}. We will use configurations of size $N_x = 64$ and $N_\tau =20$ for $\beta=7.825$ and $m_s/m_l=20$ for 2 light quarks and a strange quark. 

\begin{figure}
\centering
\includegraphics[width=0.9\textwidth]{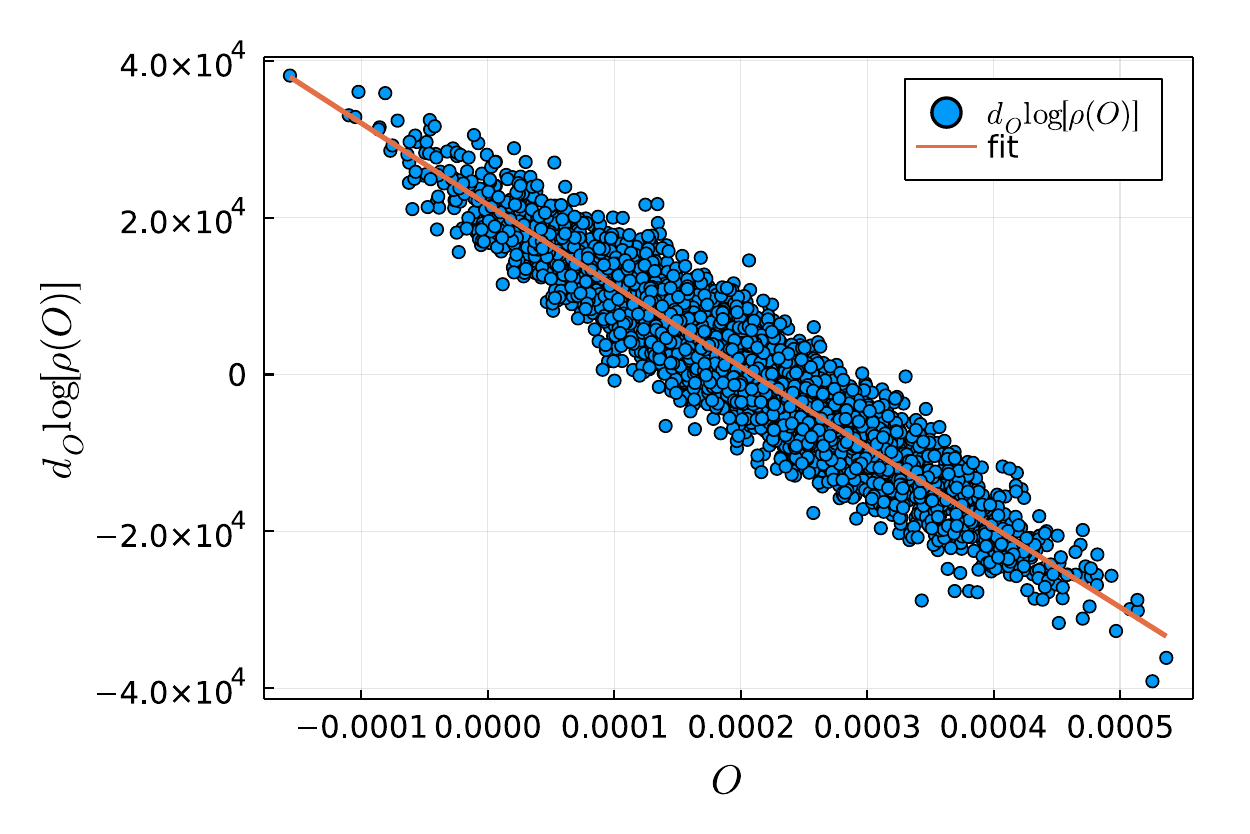}
\caption{(blue points) $4423$ samples for $d_O \log (\rho(O))$ for the Wilson line correlator O for $r/a=6$, $\tau /a=14$ and (orange line) fit done with 2. order polynomial.}
\label{fig:wilsondl}
\end{figure}

\begin{figure}
\centering
\includegraphics[width=0.9\textwidth]{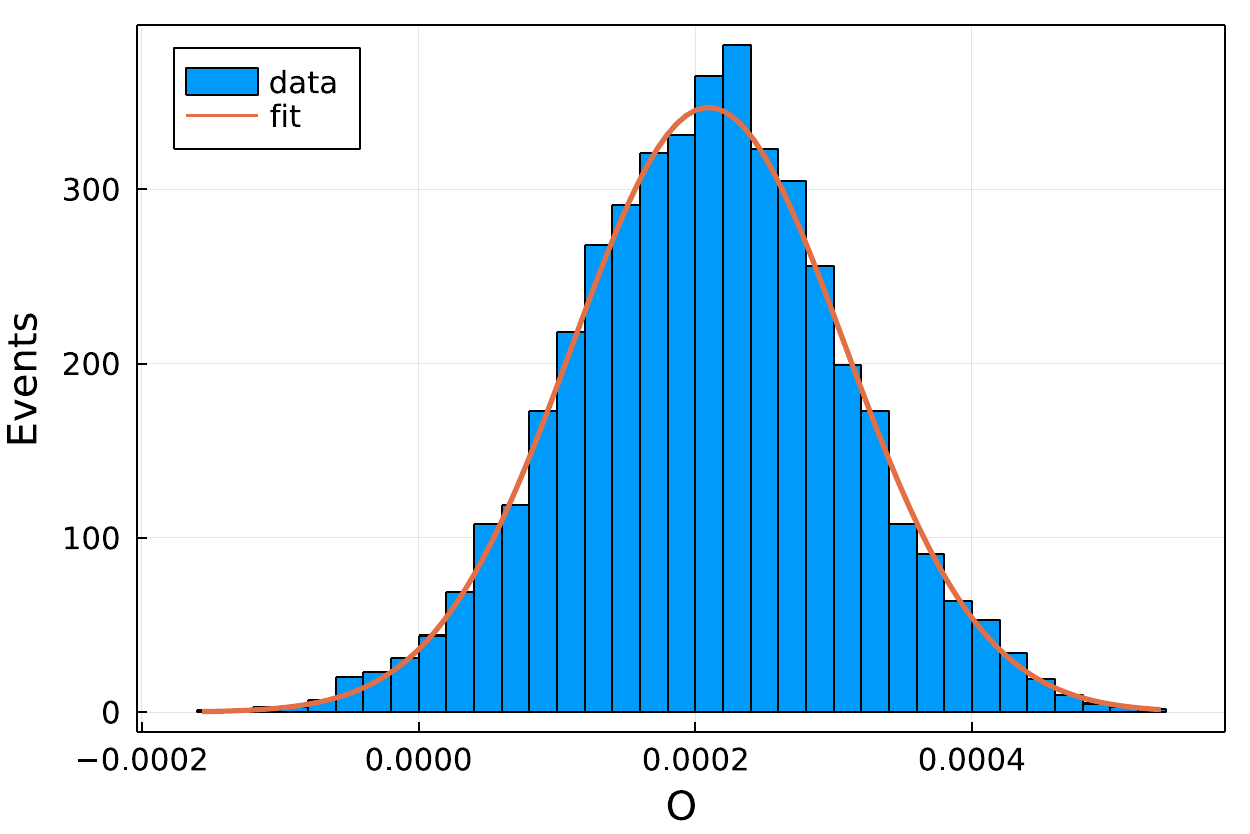}
\caption{(blue points) histogram of $4423$ samples for the Wilson line correlator O for $r/a=6$, $\tau /a=14$ and (orange line) fit done with 2. order polynomial to $\frac{d \log(\rho(O))}{d0}$ integrated up to obtain the density of states $\rho (O)$.}
\label{fig:wilsonhist}
\end{figure}

\begin{figure}
\centering
\includegraphics[width=0.99\textwidth]{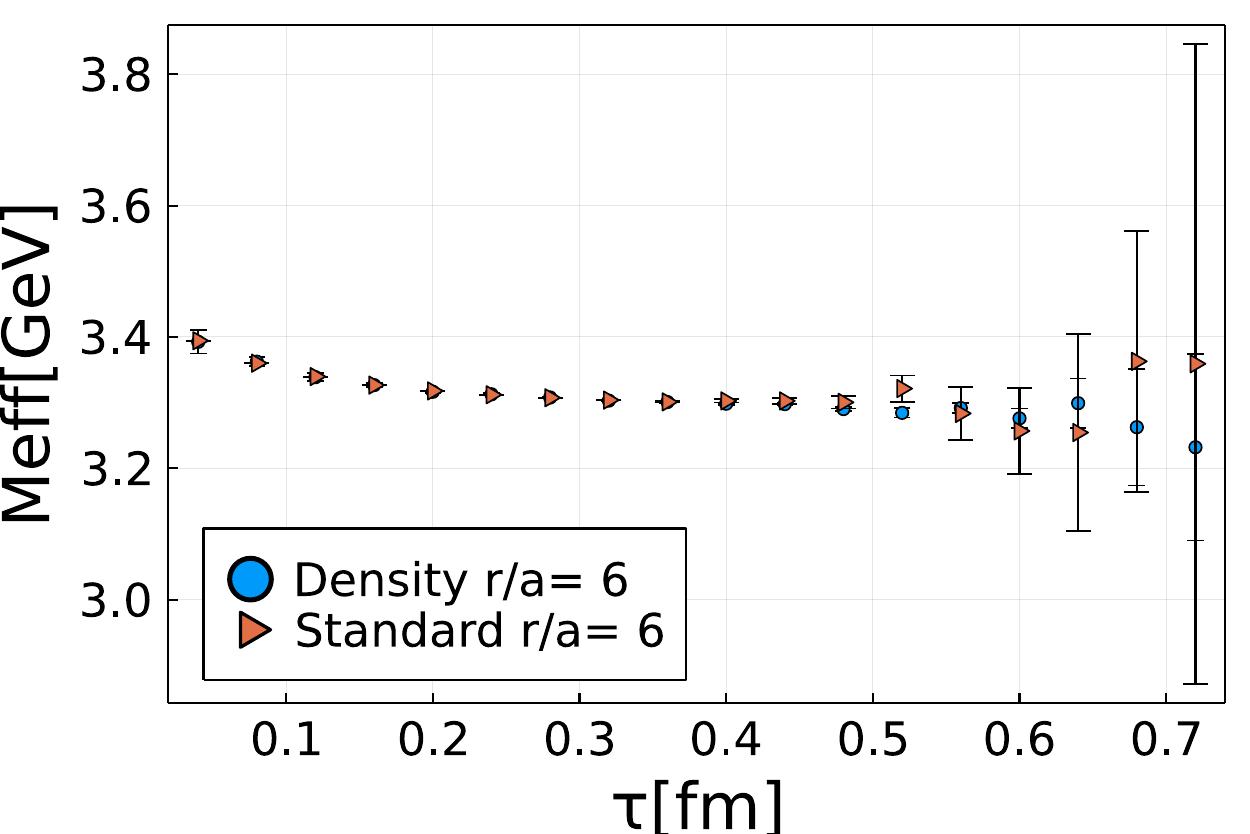}
\includegraphics[width=0.99\textwidth]{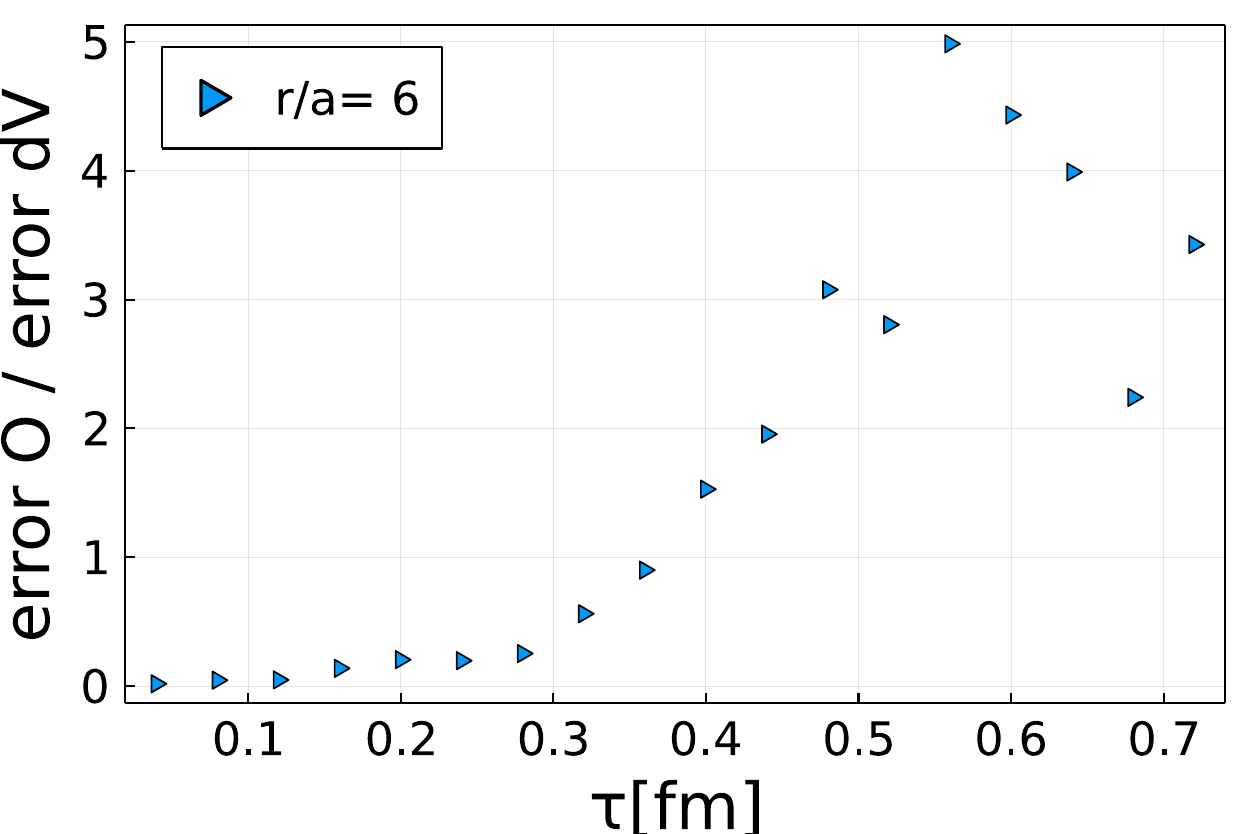}
\caption{(Top) Effective mass of the Wilson line correlator as a function of imaginary time $\tau$ at $r/a=6$ calculated using (blue) the density method $dV$ introduced in this paper and (orange) standard Monte-Carlo sampling for the observable $O$. (Bottom) Errors of standard method $O$ divided by errors from density method $dV$ for the same setup as the top plot. }
\label{fig:meffr6}
\end{figure}

\begin{figure}
\centering
\includegraphics[width=0.99\textwidth]{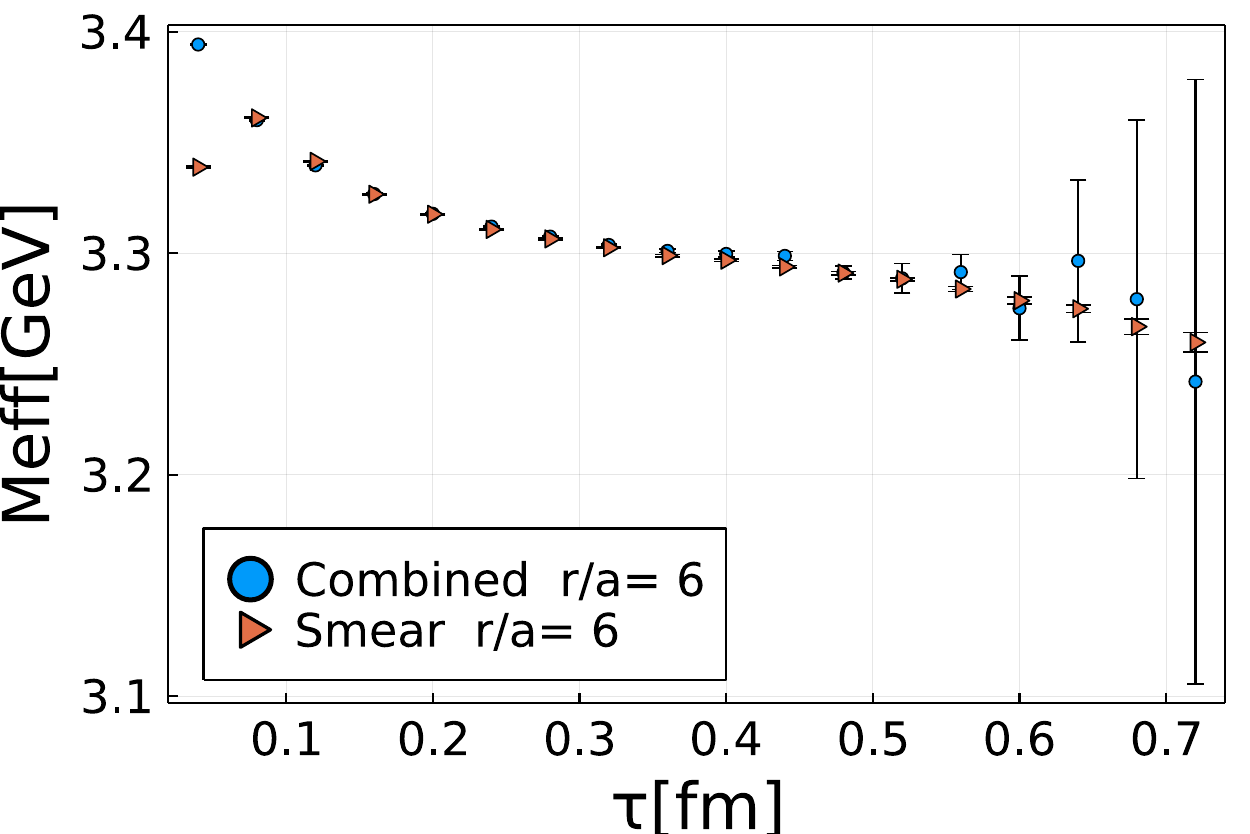}
\includegraphics[width=0.99\textwidth]{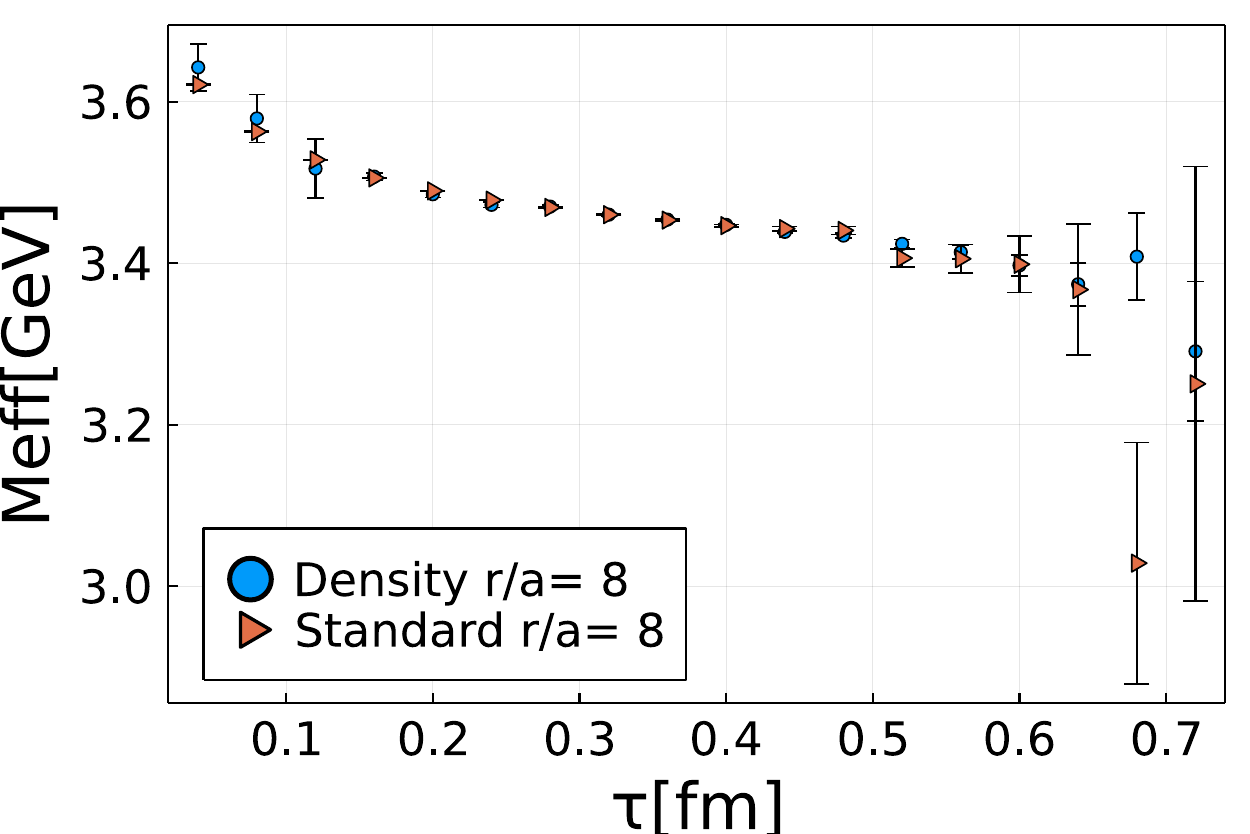}
\caption{Effective mass of the Wilson line correlator as a function of imaginary time $\tau$ at (top) $r/a=6$ and (bottom) $r/a=8$ calculated using (top blue) the density method introduced in this paper combined with the standard method, (top orange) Wilson smeared configuration, (bottom blue) density method only and (bottom orange) standard Monte-Carlo sampling.}
\label{fig:meffr8}
\end{figure}

As we sum over a large volume for each observable, the SU3 pole structure at $O={\frac{-3}{2},-1,3}$ is not seen, as the observable fluctuates at the order of $1/\sqrt{64^320}$ around the average value. The pole structure around $O=-1$ is therefore not relevant, as the results are positive and fluctuations are too small to ever get close to $O=-1$. A good fit to $\frac{d \log(\rho(O))}{d0}$ is therefore a low order polynomial, as the density of states $\rho(O)$ should be very close to a Gaussian distribution with small deviations. It was found that a second order polynomial was, for all purposes, high enough. We show the distribution and fit in figure \ref{fig:wilsondl} and \ref{fig:wilsonhist} for $\frac{d \log(\rho(O))}{d0}$ and $\rho(O)$ respectively. From the found correlation functions, we plot the effective mass
\begin{eqnarray}
M_{eff}(r,\tau) &=& \frac{1}{a}\log(C(r,\tau)/C(r,\tau +1))
\end{eqnarray}
which can be thought of as the average energy of the spectral function as a function of radius r and complex time $\tau$. We compare effective masses for the density method versus the standard method and smeared results in figure \ref{fig:meffr6} and \ref{fig:meffr8}. Smearing the gauge configurations with Wilson smearing for the smeared results does introduce artifacts at small $r$, $\tau$ and $N_\tau -\tau$, but this should only be around the first and last point in $\tau$ in the figure, and is therefore a good measurement for comparison. We also combined the standard method with the density method, assuming that they are not correlated.

At small radii, a large cancellation between the force term and the volume term has to cancel, which gives large errors due to the distribution of $\frac{d \log(\rho(O))}{d0}$ being very wide for every value of $O$, but as the complex time $\tau$ increases, the volume starts to dominate and the relative error improves as the distribution becomes thinner, until around half lattice size $N_\tau  /2 =10$, where the density method starts to be better than the standard method.

\section{Conclusion}
We have derived a formula for the local change to the density of states for a real scalar observable in eq. (\ref{eq:dO}) found from the derivatives of the observable $O$ and action $S$. From this formula, we have shown how to integrate the change to the density to obtain the density of states $\rho(O)$. First in a simple analytic case, but afterward for examples that needed $\frac{d \log(\rho(O))}{d0}$ to be sampled from a Monte-Carlo simulation. Last, we applied the method to $2+1$ lattice QCD simulations for the measurement of the Wilson line correlator. By arranging the sum of different separations with the same length, we could simplify the calculations of the density, such that the calculation was only around 4 times more expensive than the normal Wilson line correlator measurement, and a lot cheaper than generating new configurations. From this, we obtained up to a factor of 4 smaller error bars, as seen in figure \ref{fig:meffr6}. The largest improvement was at large imaginary time $\tau$, which is where it is mostly needed. The proposed method does not require new configurations to be generated, but can be used on standard Monte-Carlo samples.

The improvement from using the proposed methods depends on a couple of things. First, how expensive it is to calculate the derivative and trace of the double derivative of the observable compared to the usual method and or the time to generate the configurations. The second is a little bit harder to evaluate, but it appears that the gain is larger for less correlated samples. Essentially, if there is a large correlation, the samples between 2 uncorrelated samples are similar to estimating the slope between the 2 points, thus producing similar information. Third, the variation in $\frac{d \log(\rho(O))}{d0}$ can vary strongly. The smaller the variation, the better the precision of the obtained average.

\section*{Acknowledgements}
Rasmus Normann Larsen acknowledges support from the Research Council of Norway under the FRIPRO Young Research Talent grant 286883.
This research used awards of computer time provided by the National Energy Research Scientific Computing Center (NERSC), a U.S. Department of Energy Office of Science User Facility located at Lawrence Berkeley National Laboratory, operated under Contract No. DE-AC02- 05CH11231, and the PRACE awards on Marconi100 at CINECA, Italy.

\FloatBarrier

\begin{backmatter}




\bibliographystyle{stavanger-mathphys}

\begin{thebibliography}{9}

\bibitem{Langfeld:2012ah}
K.~Langfeld, B.~Lucini and A.~Rago,
``The density of states in gauge theories,''
Phys. Rev. Lett. \textbf{109}, 111601 (2012)
doi:10.1103/PhysRevLett.109.111601
[arXiv:1204.3243 [hep-lat]].



\bibitem{Larsen:2022pni}
R.~N.~Larsen,
``Reducing the Sign Problem with Line Integrals,''
[arXiv:2205.02257 [hep-lat]].

\bibitem{HotQCD:2023ghu}
L.~Mazur \textit{et al.} [HotQCD],
``SIMULATeQCD: A simple multi-GPU lattice code for QCD calculations,''
[arXiv:2306.01098 [hep-lat]].

\bibitem{Bazavov:2023dci}
A.~Bazavov, D.~Hoying, O.~Kaczmarek, R.~N.~Larsen, S.~Mukherjee, P.~Petreczky, A.~Rothkopf and J.~H.~Weber,
``Un-screened forces in Quark-Gluon Plasma?,''
[arXiv:2308.16587 [hep-lat]].

\bibitem{latticebook}
Jan Smith, Introduction to Quantum Fields on a Lattice,
Cambridge University Press, 2002

\end{thebibliography}


\end{backmatter}


\end{document}